\documentclass[aps,prstb,preprint,onecolumn,showpacs,groupedaddress]{revtex4-1}
\usepackage{graphicx}
\usepackage{epstopdf}
\usepackage{dcolumn}
\usepackage{bm}
\usepackage{amssymb}
\usepackage{amsmath}
\usepackage{mathalfa}
\usepackage[mathscr]{euscript}
\usepackage{relsize}
\usepackage{phonetic}
\usepackage{url}
\RequirePackage{lineno}

\usepackage{natbib}
\usepackage{har2nat}

\begin{document}

\title{$\boldmath{\check{\text{C}}}$erenkov free-electron laser with side walls}

\title{$\boldmath{\check{\text{C}}}$erenkov free-electron laser in side-wall configuration}
\author{Yashvir Kalkal}
\email{yashvirkalkal@gmail.com}
\author{Vinit Kumar}

\affiliation{Homi Bhabha National Institute, Mumbai 400094, India \\
Accelerator and Beam Physics Laboratory, Raja Ramanna Centre for Advanced Technology, Indore 452013, India}

\begin{abstract}
In this paper, we have proposed a $\check{\text{C}}$erenkov free-electron laser (CFEL) with metallic side walls, which are used to confine an electromagnetic surface mode supported by a thin dielectric slab placed on top of a conducting surface. This leads to an enhancement in coupling between the optical mode and the co-propagating electron beam, and consequently, performance of the CFEL is improved. We set up coupled Maxwell-Lorentz equations for the system, in analogy with an undulator based conventional FEL, and obtain formulas for the small-signal gain and growth rate. It is shown that small signal gain and growth rate in this configuration are larger compared to the configuration without the side walls. In the nonlinear regime, we solve the coupled Maxwell-Lorentz equations numerically and study the saturation behaviour of the system. It is found that the $\check{\text{C}}$erenkov FEL with side walls saturates quickly, and produces powerful coherent terahertz radiation.
\end{abstract}



\maketitle
\section{\label{sec:level1}Introduction}
During recent times, terahertz (THz) radiation is widely used to investigate the spectral signatures of biological and chemical molecules, and in imaging and security related applications~\cite{Siegel}. $\check{\text{C}}$erenkov free-electron laser (CFEL)~\cite{CFEL,Walsh1,Tripathi1,Garate3,Garate2,FreundGanguly,AnnularCFEL,fuente1,Fuentethesis,Kheiri,Wang,DingPlasma,DingChina,Walsh2,Walsh3,CiocciCimento,EXPCFEL2,Owens1,Owens2,Brau1,Li3}, which uses a low energy electron beam is seen as a compact source of tunable, high power coherent THz radiation. In the basic configuration of a CFEL, a dielectric slab placed on an ideal conductor supports an electromagnetic surface mode. Under suitable conditions, the electron beam propagating in the close vicinity of the dielectric surface can interact with the co-propagating surface mode to produce coherent electromagnetic radiation.

In the single slab configuration of CFEL, the surface mode is exponentially decaying in the direction perpendicular to the dielectric surface i.e., the $x$-direction; it is mainly propagating in the $z$-direction, which is the direction of the propagation of the electron beam, and only diffracts in the $y$-direction. Including the effects due to diffraction, the effective mode width in the $y$-direction can be written as $\Delta y_e = \sqrt{\pi\beta_p\lambda Z_R/4}$, where $\beta_p=v_p/c$, $v_p$ is the phase velocity of the surface mode, $c$ is the speed of light, $\lambda$ is the free space wavelength, and $Z_R$ is the Rayleigh range of the optical mode. Note that the effective mode width is here taken as $\pi$ times the rms width~\cite{KimPRSTB}. We can set $Z_R=L$, where  $L$ is length of the dielectric slab. This choice ensures that the variation in the optical beam size is within 10 $\%$ over the interaction length $L$, if we assume that the optical beam waist is at the middle of the dielectric i.e., at $z=0$. The small-signal gain of a CFEL driven by a flat electron beam is proportional to the effective surface current density $K = I/\Delta y_e$~\cite{CFEL,3D}, where $I$ is the electron beam current, and $\Delta y_e = \sqrt{\pi\beta_p\lambda L/4}$. For the low gain CFEL system, the small-signal gain has a cubic dependence on the length $L$~\cite{CFEL,3D}, and therefore, we would like to increase $L$ to increase the value of gain. However, as we increase the value of $L$ to get a reasonable value of gain, the value of $\Delta y_e$ increases, which reduces the gain. The question therefore arises whether we can make the effective mode width $\Delta y_e$ independent of $L$, and can reduce it below the value described above. It turns out that this can be achieved with the help of waveguiding, as it is done in cylindrical~\cite{Walsh1,Tripathi1,Garate3,Garate2,FreundGanguly,AnnularCFEL,fuente1,Fuentethesis,Kheiri} or rectangular~\cite{Wang,DingPlasma,DingChina} geometry of CFELs, and in conventional undulator based FELs~\cite{WaveguideFEL}. This helps in increasing the gain, and thus achieving a reasonable value of gain in a shorter interaction length.

In the cylindrical waveguide geometry of CFELs, a metallic waveguide of cylindrical shape is lined with a dielectric material on its inner surface~\cite{Walsh1,Tripathi1,Garate3,Garate2,FreundGanguly,AnnularCFEL,fuente1,Fuentethesis,Kheiri}, and in the rectangular waveguide geometry of CFELs, a rectangular metallic waveguide is used, with a dielectric layer on the bottom surface~\cite{DingPlasma,DingChina}. Both the cylindrical and the rectangular waveguide geometries of CFELs are closed structures. Over the years, many experimental~\cite{Walsh1,AnnularCFEL,Fuentethesis,Wang} and theoretical~\cite{Tripathi1,Garate3,Garate2,FreundGanguly,fuente1,Kheiri,DingPlasma,DingChina} studies have been performed on the guided surface modes supported in these configurations. Walsh $et~al.$~\cite{Walsh1} determined the dispersion relation of the cylindrical geometry based CFEL by treating the electron beam as a linear fluid, and subsequently the growth rate was evaluated by performing the Taylor series expansion of the dispersion relation around the roots of no-beam dispersion~\cite{Garate3}. The instability of cylindrical geometry based CFELs in the linear regime was also studied by setting up the coupled Maxwell-Vlasov equations~\cite{Tripathi1,Garate2,Kheiri}. The three-dimensional (3D) non-linear analysis of CFELs in the cylindrical geometry was performed by setting up the first order coupled Maxwell-Lorentz equations by Freund and Ganguly~\cite{FreundGanguly}. Fuente $et~al.$~\cite{fuente1,Fuentethesis} extended this approach to include the effect of liner fluctuations and dielectric losses on the performance of a cylindrical geometry based CFEL system. It is important to note that due to the presence of dielectric liner, the structure supports electromagnetic waves with phase velocity less than the speed of light, such that the electron beam can interact with the co-propagating electromagnetic wave. The electromagnetic field for such modes is confined within the liner and decays, when we move away from the liner and approach towards the axis of the waveguide. The decay rate will be very small for the electromagnetic wave having phase velocity nearly equal to the speed of light. This however requires a relativistic electron beam, which can co-propagate with the electromagnetic wave and drive the CFEL system. A more practical case of interest is that when the phase velocity of the electromagnetic wave is reasonably lower than the speed of light such that a low energy electron beam can exchange energy with the co-propagating electromagnetic wave and drive the CFEL system. However, in this case, the field decays faster, while moving away from the dielectric liner, in the vacuum region. In such a situation, it becomes advantageous to use a hollow electron beam having a radius nearly equal to the radius of the waveguide to have an effective beam-wave interaction. For operation at higher frequencies, the transverse dimension of cylindrical waveguide is required to be small~\cite{Fuentethesis}, and we therefore need to use a hollow electron beam of very small radius. This will reduce the cross sectional area of the beam, and therefore will increase the space charge effects, which may inhibit the high power operation of the device. Also, in order to ensure that the beam remains close to the dielectric surface as it propagates, the hollow cylindrical beam is required to have very stringent transverse emittance in the radial direction, which may be difficult to achieve. These problems can be avoided by introducing a planar configuration i.e., CFEL in single or double slab configuration~\cite{CFEL,Walsh2,Walsh3,CiocciCimento,EXPCFEL2,Owens1,Owens2,Brau1,Li3}, and CFEL in rectangular waveguide geometry~\cite{DingPlasma,DingChina}. In the planar configuration, we can use a flat electron beam, which remains close to the dielectric surface, but has relatively larger size in the horizontal direction. Here, the cross sectional area of the beam can be larger such that the space charge effect is relatively reduced, and also we need to maintain the stringent emittance only in the vertical direction, and not in the horizontal direction. Ding $et~al.$~\cite{DingPlasma,DingChina} have developed a 3D approach similar to the given in Ref.~\cite{FreundGanguly} for the case of CFELs in the rectangular waveguide geometry; and also included the second and higher order variations of wave amplitude. In the rectangular waveguide geometry of CFEL, the field decays in the vertical direction as we move away from the dielectric surface. Although the surface mode is already confined close to the dielectric surface, the top surface of the rectangular waveguide helps in further confining the electromagnetic surface mode if the vertical dimension of the waveguide is small. This however results in attenuation of the wave due to heat dissipation on the top surface, and also makes the structure a closed one. One option is to remove the top surface, such that the structure is \textit{open}, and thus uses only two side walls to confine the surface mode in the horizontal direction. In this paper, we study this configuration, which is shown in Fig.~1, and present a detailed analysis, including the effect of finite size, energy-spread and emittance of the electron beam; and also the attenuation of the surface mode.

Due to the presence of metallic side walls, the surface mode is guided in the $y$-direction, and maintains a constant width over any arbitrary length. Smaller value of mode width results in a good overlapping between the electron beam and the copropagating guided surface mode; and consequently the gain of CFEL can be increased, which is the primary advantage here. The second advantage is that the requirement on vertical emittance of the electron beam is relaxed with the help of side walls. This can be understood as follows. The surface mode supported in the CFELs is evanescent in the direction perpendicular to the dielectric surface with scale height $h=\beta_p\gamma_p\lambda/4\pi$~\cite{CiocciCimento}, where $\gamma_p=1/\sqrt{1-\beta_p^2}$. To maintain a good overlap with the radiation beam, the electron beam has to maintain its vertical beam size around this value over the entire interaction length. As discussed in Refs.~\cite{CiocciCimento,EXPCFEL2}, one of the challenges in a CFEL is to maintain a very small vertical beam size over the entire interaction length. With the help of waveguiding, the required interaction length is reduced, and therefore, we need to maintain a small vertical beam size now over a \textit{smaller} interaction length, which relaxes the requirement on the vertical beam emittance. The third advantage of the sidewall configuration is that it helps in reducing the loss due to the attenuation of surface mode, which is caused by the finite conductivity of metal, and the dielectric loss. For a low gain CFEL oscillator system, the small-signal gain has cubic dependence on $L$, whereas the attenuation results in exponential decay of power by $e^{-4\alpha L}$ for a round trip, where $\alpha$ is the field attenuation coefficient. In order to reduce the degradation in net gain due to attenuation, it is therefore desired to reduce the interaction length. With the help of waveguiding, we can take a shorter interaction length, and still obtain higher gain such that the device can produce powerful electromagnetic radiation. We would like to add that the concept of side walls has been introduced earlier in the Smith-Purcell free-electron lasers (SP-FELs)~\cite{SWLi,SWAJB,SWDonohue}, and sidewall grating structure is found to be advantageous compared to the general grating (without side walls). We expect similar improvement for the CFELs.

\begin{figure}
\begin{center}
\includegraphics[width=12.0 cm,height=5.5 cm]{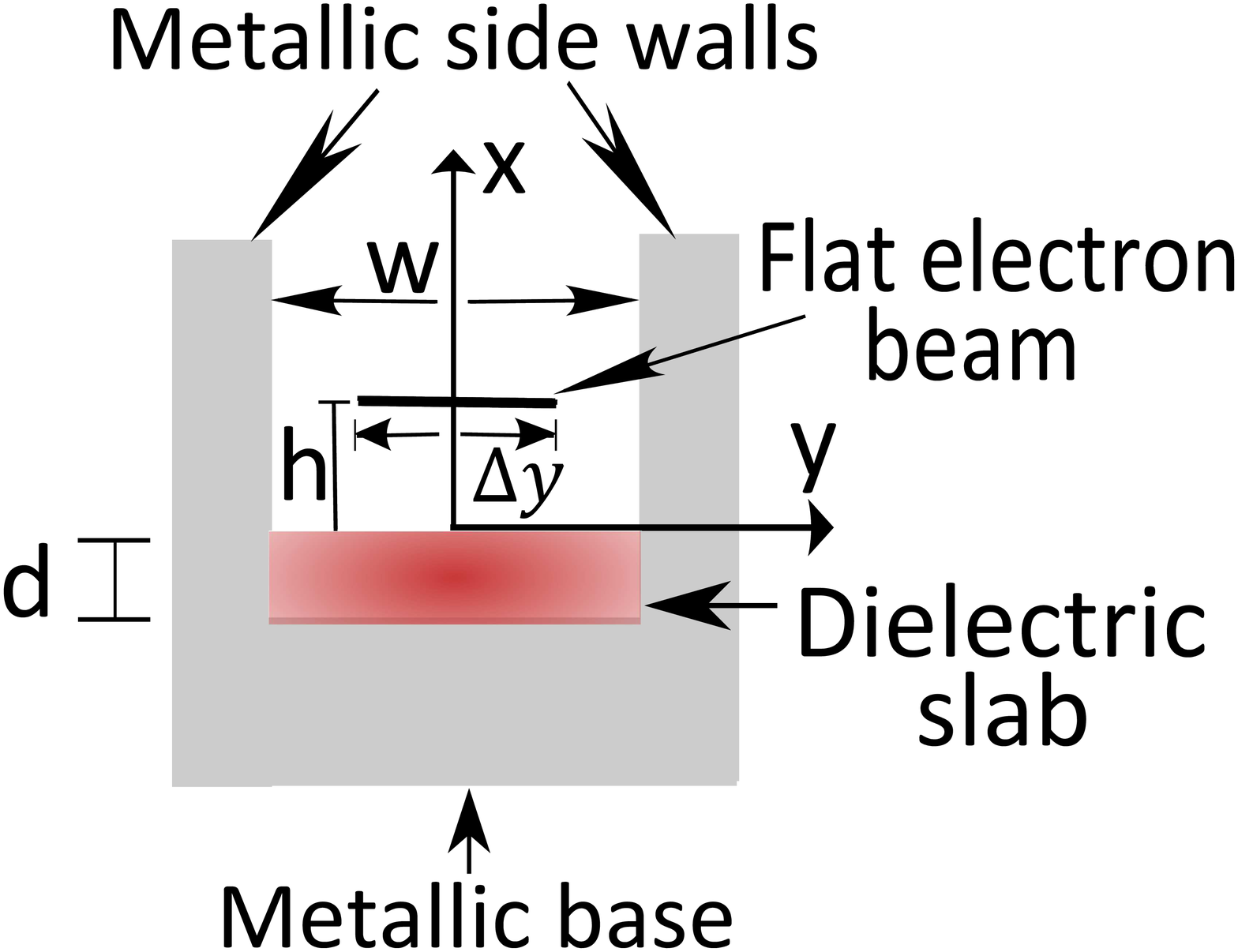}
\end{center}
\caption{Schematic of a $\check{\text{C}}$erenkov FEL with metallic side walls.}
\end{figure}

In this paper, we present a detailed analysis of single slab based CFEL in the sidewall configuration. In the next section, we will derive the dispersion relation of the surface mode supported by the CFEL in the sidewall configuration, and derive the coupled Maxwell-Lorentz equations to describe the interaction between the surface mode and the flat electron beam. We will then discuss an approximate analytical solution of these equations in the linear regime for the small-signal gain and the growth rate, and also discuss the results of the numerical simulations performed to solve the coupled Maxwell-Lorentz equations in the non-linear regime, in the same section. The calculations for the power, group velocity, and the attenuation coefficient of the surface mode supported in a sidewall CFEL are discussed in Appendix~A. In Appendix~B, we set up the coupled Maxwell-Lorentz equations for a finite-thickness beam.

\section{\label{sec:level2}Theory and calculations}
\subsection{Formula for the resonant wavelength}

The schematic of $\check{\text{C}}$erenkov FEL with side walls is shown in Fig. 1, where a dielectric slab with thickness $d$, length $L$, and relative dielectric permittivity $\epsilon$ is placed on the metallic surface. In the $y$-direction, the dielectric slab is surrounded with metallic side walls with spacing $w$. For simplicity, we will first perform the analysis for a flat electron beam having vanishing thickness in the $x$-direction, and having width $\Delta y$ in the $y$-direction. The flat electron beam is here assumed to be propagating with velocity $v$ along the $z$-direction, at a height $h$ above the dielectric surface. We would like to clarify here that the flat beam is actually a simplified way of representing the beam having a thickness $2h$ in the $x$-direction, with its centroid at a height $h$ above the dielectric surface. The thickness, as well as the emittance of the beam in the $x$-direction is much smaller compared to the corresponding value in the $y$-direction. Later, we will generalize our results by explicitly taking into account the finite-thickness of the electron beam in the $x$-direction.

The electromagnetic surface mode supported by this geometry can be obtained by combining the plane wave solutions of an open structure, i.e., the structure without any side wall, in a suitable manner such that it satisfies the boundary conditions. Field components of the lowest order TM surface mode, which satisfy the Maxwell equations with the given boundary conditions are discussed in detail in Appendix~A. The expression for the longitudinal electric field $E_z^I$ is given by,
\begin{eqnarray}
E_z^I &=& E_0 e^{-\Gamma(x-h)}e^{i(k_z z -\omega t)}\cos(k_y y)+\text{c.c.,}~~~~~~~~~~~~~~~~~\nonumber\\ 
&=&\frac{E_0}{2}\big[e^{i(k_z z+k_yy -\omega t)}+e^{i(k_z z-k_yy -\omega t)}\big]e^{-\Gamma(x-h)}+\text{c.c..}
\end{eqnarray}
Here, $2E_o$ is defined as the peak amplitude of $E_z^I$ at the location of the electron beam at $x=h$, $k_z=2\pi/\beta \lambda$ is the propagation wavenumber in the $z$-direction, $\beta=v/c$, $\omega=2\pi c/\lambda$,  $\lambda$ is the free space wavelength, $\Gamma=\sqrt{k_y^2+k_z^2-\omega^2/c^2}$, $k_y=\pi/w$, and c.c. denotes complex conjugate. The phase velocity of the surface mode is here taken as equal to the electron velocity $v$. As it is seen in the above equation, the guided surface mode is a combination of two plane evanescent waves travelling in different directions in the ($y,z$) plane, each having frequency $\omega$ and wave vector $k_0 = \sqrt{k_y^2+k_z^2}$. Each of these two plane evanescent waves is a solution of the wave equation for the case of CFEL without side walls, and  should therefore satisfy the corresponding dispersion relation $k_0=\tan^{-1}(1/a)/b$ for that case~\cite{CFEL,Walsh2,Owens1}. Here, $a=(\gamma_p/\epsilon)\sqrt{\epsilon\beta_p^2-1}$, $b=d\sqrt{\epsilon\beta_p^2-1}$, and $\beta_p = \omega/{ck_0}$. Note that here we have used the property that the dielectric slab is an isotropic structure in the ($y,z$) plane. Using this dispersion relation, we obtain the following expression for the operating wavelength of the sidewall CFEL:
\begin{eqnarray}
\lambda=\frac{2\pi}{\beta \sqrt{[\tan^{-1}(1/a)/b]^2-[\pi/w]^2}}.
\end{eqnarray}
\begin{figure}
\includegraphics[width=16.0cm]{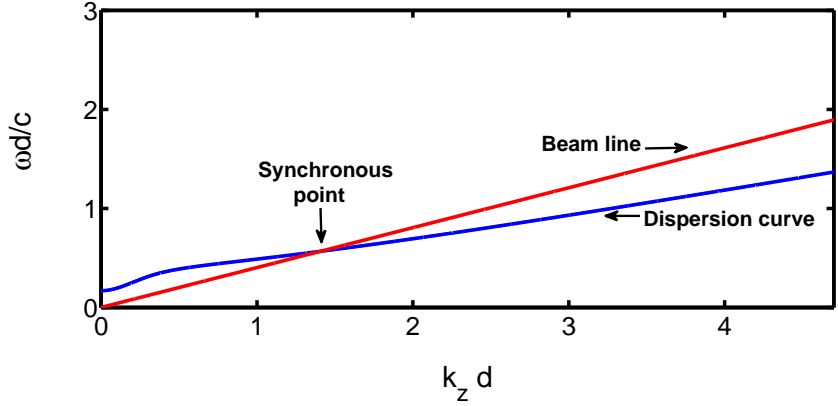}
\caption{Plot of the dispersion curve of the electromagnetic surface mode (in the empty structure, i.e., without the electron beam), and the Doppler line for the electron beam. The resonant frequency of the CFEL system is obtained at the intersection.}
\end{figure} 
From the above equation, it is clear that for a finite sidewall spacing $w$, we typically require an electron beam with higher energy to achieve the same operating wavelength, as compared to the case of a CFEL without any side wall, provided that all other parameters are same. The dispersion curve for the CFEL with side walls is shown in Fig.~2. The parameters used in our calculations are summarized in Table 1. For the dielectric slab, we choose GaAs material, which is an isotropic material with relative dielectric permittivity $\epsilon = 13.1$~\cite{Owens2}. As discussed in detail in the next section, we have taken side wall spacing $w=(2/3)\sqrt{\pi\beta_p\lambda L/4}$, which gives an enhancement in small-signal gain up to a factor of three, compared to the case of CFEL without any side wall. The resonant frequency is obtained as 0.11 THz for these parameters. 
\begin{table}[b]
\caption{\label{tab:table1}Parameters of a sidewall CFEL used in the calculation}
\begin{ruledtabular}
\begin{tabular}{lcdr}
~~~~~~Electron energy & 47.5 keV\\
~~~~~~Electron-beam height ($h$) & 90.5 $\mu$m\\
~~~~~~Electron-beam current ($I$) & 35 mA\\
~~~~~~Dielectric constant ($\epsilon$) &  13.1\\
~~~~~~Length of slab ($L$) & 5 cm\\
~~~~~~Dielectric thickness ($d$) & 235 $\mu$m\\
~~~~~~Side walls separation ($w$) & 4.3 mm \\
~~~~~~Operating frequency & 0.11 THz\\
\end{tabular}
\end{ruledtabular}
\end{table}

It is important to mention here that the electromagnetic surface mode given by Eq.~(1) has $\cos(k_yy)$ variation along the $y$-direction, where $k_y=\pi/w$ for the fundamental mode, and will have maximum amplitude at the middle ($y=0$) of the dielectric. Due to this, there will be maximum overlap of the surface mode with the electron beam propagating between the side walls, resulting in an effective exchange of energy. Higher order modes corresponding to $k_y = (2n+1)\pi/w$ will not have good overlap with the electron beam, and will not be able to exchange energy with the electron beam in an effective manner. We therefore do not consider them in our analysis.
\subsection{Coupled Maxwell-Lorentz equations and gain calculation}
Next, we will derive the coupled Maxwell-Lorentz equations to study the interaction of the guided surface mode with the co-propagating electron beam in the sidewall CFEL. For this, we follow an approach, which is familiar for the case of conventional FELs~\cite{ColsonFEL1}, and backward wave oscillators (BWOs)~\cite{LevushBWO}, which is a slow wave structure like CFEL. In the presented model, an ensemble of electrons interacts with the co-propagating surface mode, and we have assumed that the amplitude of the surface mode is a slowly varying function of $z$ and $t$, due to interaction with the co-propagating electron beam. Now, the electromagnetic surface field described earlier in the section can be written in a more general form as
\begin{eqnarray}
\boldsymbol{E^T}&=& \big[{A(z,t)}\boldsymbol{E_p}(x,k_y)+\boldsymbol{E_{sc}}\big]e^{i(k_zz-\omega t)}+\text{c.c.,}\\
\boldsymbol{B^T}&=& \big[{A(z,t)}\boldsymbol{B_p}(x,k_y)+\boldsymbol{B_{sc}}\big]e^{i(k_zz-\omega t)}+\text{c.c.},~
\end{eqnarray} 
where, $A(z,t)$ is the amplitude of the fundamental mode, and the symbols $\boldsymbol{E_p}$ and $\boldsymbol{B_p}$ represent the field distributions for the fundamental mode in the empty structure, i.e., in the absence of electron beam. The $x,y$ and $z$ components of $\boldsymbol{E_p}$ and $\boldsymbol{B_p}$ are written explicitly in Appendix~A. Here, $\boldsymbol{E_{sc}}$ and $\boldsymbol{B_{sc}}$ denote the small first-order ac space charge  fields. We would like to mention here that we have not considered the effect of DC fields (zeroth order) since the beam is not space charge dominated. This is because for the typical beam parameters considered later in this paper, the space charge term is not very significant compared to the emittance term in the envelope equation~\cite{Chaobook} that describes the evolution of the beam. The typical value of electron beam current density is around 0.1 A/mm$^2$ for the calculations presented in this paper.

Total electromagnetic fields represented by Eqs.~(3,4) will satisfy Maxwell equations with the beam current density $\boldsymbol{J}$ to give:
\begin{eqnarray}
\frac{1}{c^2}\frac{\partial A}{\partial t} \boldsymbol{E_p} - \frac{i\omega}{c^2}\boldsymbol{E_{sc}}= \nabla\times\boldsymbol{B_{sc}} + ik_z\hat{z}\times\boldsymbol{B_{sc}} ~~~~~~\nonumber\\
\hspace*{3pt} + \frac{\partial A}{\partial z}\hat{z}\times\boldsymbol{B_p}-\mu_0\boldsymbol{J}e^{-i(k_zz-\omega t)},~~~~~~\\
-\frac{\partial A}{\partial t} \boldsymbol{B_p} +i\omega\boldsymbol{B_{sc}}= \nabla\times\boldsymbol{E_{sc}} + ik_z\hat{z}\times\boldsymbol{E_{sc}} ~~~~~~~\nonumber\\
\hspace*{3pt} + \frac{\partial A}{\partial z}\hat{z}\times\boldsymbol{E_p},~~~~~~~~~~~~~~~~~~~~~
\end{eqnarray}
where, $\mu_0$ is the permeability of the free space. By taking the dot product of Eq.~(5) with $\boldsymbol{E_p}^*$, and Eq.~(6) with $\boldsymbol{B_p}^*$, and subtracting the resultants, respectively; we obtain: 
\begin{eqnarray}
\bigg[\frac{\vert\boldsymbol{E_p}\vert^2}{c^2}+\vert\boldsymbol{B_p}\vert^2\bigg]\frac{\partial A}{\partial t}+\hat{z}.\big[\boldsymbol{E_{p}^*}\times\boldsymbol{B_{p}}+\boldsymbol{E_{p}}\times\boldsymbol{B_{p}^*}\big]\frac{\partial A}{\partial z}
=-\mu_0\boldsymbol{J}.\boldsymbol{E_p^*}e^{-i(k_zz-\omega t)}\nonumber\\
+\hspace*{3pt}\nabla.[\boldsymbol{B_{sc}}\times\boldsymbol{E_p^*}-\boldsymbol{E_{sc}}\times\boldsymbol{B_{p}^*}].
\end{eqnarray}
Note that while deriving the above equation, we have used the fact that the complex conjugate fields $\boldsymbol{E_p^*}$ and $\boldsymbol{B_p^*}$ satisfy the Maxwell wave equation in the empty structure (i.e., without the electron beam). We performed integration on both sides of Eq.~(7) over a volume containing one period of evanescent wave i.e., $\lambda_z=\beta\lambda$. The integration over $x$ is in the range $(-d,\infty)$, over $y$ is in the range $(-w/2, w/2)$, and over $z$ is in the range $(z-\lambda_z/2, z+\lambda_z/2)$. In doing this, we can assume that $A(z,t)$ varies slowly in the longitudinal direction, and can therefore be taken out of the integral. Further, the tangential components of $\boldsymbol{E_p}$ vanish at the metallic surfaces located at $x=-d$, and at $y=\pm w/2$; the same condition will be satisfied by the tangential components of $\boldsymbol{E_{sc}}$ in the presence of the electron beam. The electromagnetic fields are evanescent in the $x$-direction, and vanish at $x=\infty$. Due to these conditions, the last term on the right hand side of Eq.~(7) will vanish upon integration, and we obtain:
\begin{eqnarray}
\frac{\partial A}{\partial t}+v_g\frac{\partial A}{\partial z}=\frac{-\vert A\vert^2}{w\lambda_z\mathcal{U}} \int_{z-\lambda_z/2}^{z+\lambda_z/2}\int_{-w/2}^{w/2}\int_{-d}^{\infty}\boldsymbol{J}.\boldsymbol{E_p^*}e^{-i(k_zz-\omega t)}dx dy dz,
\end{eqnarray}
where, $v_g$ is group velocity of the surface mode, which is equal to $P/w\mathcal{U}$ for the CFEL~\cite{CFEL}, $P$ is total power contained in the surface mode, and $\mathcal{U}$ is electromagnetic energy stored in the fields per unit mode width $w$, and per unit length in the $z$-direction. The current density of the flat beam propagating at a height $h$ from the dielectric surface is given by $\boldsymbol{J}=e\sum\limits_{i} \delta (x-h)\delta(y-y_i)\delta[z-z_i(t)]v \hat{z}$, where $e$ represents the magnitude of the electron's charge, $y_i$ and $z_i$ are the coordinates of the $i$th particle in the $y$ and $z$-direction, respectively at time $t$. The $z$ component of the electromagnetic field $\boldsymbol{E_p}$, which interacts with the electron beam is given by $E_0\cos(k_yy)e^{-\Gamma (x-h)}\hat{z}$, and the term $AE_o$ is represented by $E_z$ in further calculations. Substituting $\boldsymbol{E_p^*}$ and $\boldsymbol{J}$ in Eq.~(8), and performing the integral, we obtain the following time dependent differential equation for $E_z$:
\begin{eqnarray}
\frac{\partial E_z}{\partial z}+\frac{1}{v_g}\frac{\partial E_z}{\partial t}=-\frac{IE_z^2}{v_g w\mathcal{U}}\langle\cos(k_yy) e^{-i\psi} \rangle.
\end{eqnarray} 
Here, $I=evN_{\lambda_z}/\lambda_z$ is the electron beam current, $N_{\lambda_z}$ is the number of electrons distributed over one spatial wavelength of the evanescent wave, i.e., $\lambda_z$, $\psi=k_zz-\omega t$ is the phase of the electron, and $\langle \cdots \rangle$ represents the averaging over the total number of electrons distributed over $\lambda_z$ and over the beam width $\Delta y$ in the $y$-direction. Due to interaction between the surface mode and the electron beam, electrons get bunched at the resonant frequency $\omega$ of the surface mode. The term $\langle \cos(k_yy)e^{-i\psi} \rangle$ represents weighted bunching factor, which arises due to interaction of the electron beam with the co-propagating surface mode. Next, we take the effect of attenuation of the surface mode due to the ohmic losses present on the surface of conductor, and due to the losses present in the dielectric medium. In the presence of losses, the surface wave will attenuate as it propagates, and by including the effect of attenuation, we write the following generalized time dependent differential equation for the $E_z$:
\begin{eqnarray}
\frac{\partial E_z}{\partial z}+\frac{1}{v_g}\frac{\partial E_z}{\partial t}=-\frac{IE_z^2}{v_g w\mathcal{U}}\langle\cos(k_yy) e^{-i\psi} \rangle-\alpha E_z.
\end{eqnarray} 
Here, $\alpha$ is the field attenuation coefficient. The detailed calculation for $\alpha$ is discussed in Appendix~A.

Now, we discuss the longitudinal dynamics of the electron beam. We neglect the transverse motion of the electrons, and write the equations for the evolution of energy and phase of $i$th electron respectively as~\cite{CFEL}:
\begin{eqnarray}
\frac{\partial \gamma_i}{\partial z}+\frac{1}{v}\frac{\partial \gamma_i}{\partial t}=\frac{eE_z}{mc^2}\cos(k_yy)e^{i\psi_i}+\text{c.c.,}~~~~~~~~
\end{eqnarray}
\begin{eqnarray}
\frac{\partial \psi_i}{\partial z}+\frac{1}{v}\frac{\partial \psi_i}{\partial t} =\frac{\omega}{c\beta_R^3\gamma_R^3} (\gamma_i-\gamma_R).~~~~~~~~~~~~~~~~~
\end{eqnarray}
Here, subscript $i$ is meant for the $i$th particle, subscript $R$ is meant for the resonant particle, $\beta_R=v_R/c$, $\gamma_R=1/\sqrt{1-\beta_R^2}$ is the relativistic Lorentz factor, and $m$ is mass of the electron. At resonance, the electron velocity $v$ is equal to the phase velocity of the co-propagating evanescent surface mode along the $z$-axis. Note that the electromagnetic field has $\cos(k_yy)$ type variation and it has peak value at $y=0$. Electrons will see the maximum field at $y=0$ while propagating along the $z$-direction. For the parameters of CFEL discussed earlier, the ac space charge field does not have any significant effect on the dynamics of the electron beam~\cite{CFEL}, and we have therefore neglected it in our calculations. Equations (10-12) can be expressed in a more elegant form by defining the following dimensionless variables~\cite{CFEL,VinitPRE}:
\begin{equation}
\xi = z/L,
\end{equation}
\begin{eqnarray}
\tau=\bigg( t -\frac{z}{v _R}\bigg) \bigg(\frac{1}{v_g}-\frac{1}{v_R} \bigg)^{-1} \frac{1}{L},
\end{eqnarray}
\begin{equation}
\eta_i=\frac{k_zL}{\beta_R^2\gamma_R^3}(\gamma_i-\gamma_R),
\end{equation}
\begin{equation}
\mathcal{E}=\frac{4\pi k_zL^2}{I_AZ_0\beta_R^2\gamma_R^3}E_z,
\end{equation}
\begin{equation}
\mathcal{J}= \frac{4\pi k_zL^3}{Z_0\beta_R^2\gamma_R^3}\frac{I}{I_A}\frac{E_z^2}{v_g w\mathcal{U}}.
\end{equation}
Here, $\xi$ is the dimensionless distance, which varies from 0 to 1, and $\tau$ is the dimensionless time variable, having an offset of $z/v_R$ from the real time $t$, $\mathcal{E}$ is the dimensionless surface mode field, and the normalized energy detuning of the $i$th electron is $\eta_i$. The dimensionless beam current is written as $\mathcal{J}$, $Z_0=1/\epsilon_0c=377~\Omega$ is the characteristic impedance of the free space, $\epsilon_0$ is absolute permittivity of the free space, and $I_A=4\pi\epsilon_0mc^3/e=17.04$ kA is the Alfv$\acute{\text{e}}$n current. With these dimensionless variables, the set of Eqs.~(10-12) assumes the following form:
 \begin{equation}
 \frac{\partial \mathcal{E}}{\partial \xi}+\frac{\partial \mathcal{E}}{\partial \tau}= -\mathcal{J}\langle\cos(k_yy)e^{-i\psi}\rangle-\alpha L \mathcal{E},~~~~~~
 \end{equation}
 \begin{equation}
 \frac{\partial \eta_i}{\partial \xi}=\mathcal{E}\cos(k_yy)e^{i\psi_i}+\text{c.c.,}
 \end{equation}
  \begin{equation}
  \frac{\partial \psi_i}{\partial \xi}=\eta_i.~~~~~~~~
  \end{equation}
Equations (18-20) are known as the coupled Maxwell-Lorentz equations, and these equations have to be solved numerically to study the detailed behaviour of the CFEL system with side walls. In the limit of narrow electron beam i.e., $\Delta y\ll w$, and in the small-signal small-gain regime, we can find an approximate analytical solution of these equations by following a procedure given in Ref.~\cite{Braubook}. Neglecting the attenuation effect, and taking the initial detuning parameter $\eta=2.6$ to maximize the gain~\cite{Braubook}, we obtain the expression for the small-signal gain in a single pass operation of the sidewall CFEL as:
\begin{eqnarray}
G= 6.75\times 10^{-2} \frac{16\pi k_zL^3}{Z_0\beta_R^2\gamma_R^3}\frac{I}{I_A}\frac{E_z^2}{v_g w\mathcal{U}}\bigg(\frac{\sin(k_y\Delta y/2)}{k_y\Delta y/2}\bigg)^2.
\end{eqnarray} 
The expression for $\mathcal{U}/E_z^2$ is given by Eq.~(A.12) in Appendix~A, through which it can be seen that the gain has $e^{-2\Gamma h}$ dependency on the height $h$ of the flat electron beam. Note that, as we move away from the limit $\Delta y\ll w$, the value of gain obtained by numerically solving the Eqs.~(18-20) may no longer be in good agreement with that obtained using Eq.~(21). We observe that the above expression is similar to the expression for the gain derived for the case of CFEL without any side wall in Ref.~\cite{CFEL}, with some notable differences. First, the parameter $w$ appearing in the denominator of the above equation appears as $\Delta y_e=\sqrt{\pi\beta_p\lambda L/4}$ in the corresponding equation for the case of CFEL without any side wall. Here, $w$ is independent of $L$ because of waveguiding, and can be made much smaller compared to $\Delta y_e$. Next, by comparing the expression for $\mathcal{U}/E_z^2$ given by Eq.~(A.12), with the corresponding equation given in Ref.~\cite{CFEL} for the case of CFEL without any side wall, we observe that the value of $\mathcal{U}/E_z^2$ is reduced nearly by a factor of 2 due to the presence of side walls. Thus the effective mode width for the case of CFEL with side walls can be taken as $w/2$. Hence, if we choose $w=(2/3)\Delta y_e=(2/3)\sqrt{\pi\beta_p\lambda L/4}$, we expect an enhancement in gain up to a factor of 3. Finally, we observe that there is a term containing the square of a sinc  function on the right hand side of Eq.~(21). This term is maximum if $\Delta y \rightarrow 0$, which means that if all the electrons are at $y=0$, they experience the peak of the electric field amplitude, and the gain is maximum. Taking the effect of attenuation, there will be a single pass loss ($1-e^{-2\alpha L}$) in addition to the gain given by Eq.~(21).

In the small-signal high-gain regime, and for limit $\Delta y\ll w$; we have solved coupled Maxwell-Lorentz equations by using the collective variables as described in Ref.~\cite{CFEL}, and found the growth rate of the system as:
\begin{eqnarray}
\mu=\frac{\sqrt{3}}{2L}\bigg[ \frac{4\pi k_zL^3}{Z_0\beta_R^2\gamma_R^3}\frac{I}{I_A}\frac{E_z^2}{v_g w\mathcal{U}}\bigg(\frac{\sin(k_y\Delta y/2)}{k_y\Delta y/2}\bigg)^2\bigg]^{1/3}.
\end{eqnarray}
Note that the above equation is derived in the absence of the attenuation of the surface wave. Due to the effect of attenuation, the net growth rate becomes $\mu-\alpha$.

\subsection{Numerical simulations}

To study the saturation behaviour of the system, we have solved the coupled Maxwell-Lorentz equations numerically by writing a computer code using the Leapfrog method~\cite{Leapfrogbook}. We will seek for the steady state solutions of Eqs.~(18-20). The input electron beam is taken as a DC beam. In order to evaluate the first term on the right side of Eq.~(18), we need to perform an averaging over the electrons distributed over one wavelength of the evanescent wave. Hence, in the simulation, we need to take the number of electrons same as the number of electrons distributed over one spatial wavelength of the evanescent wave i.e., $N_{\lambda_z}=I\lambda_z/ev_R$. We obtain $N_{\lambda_z}\simeq 2^{21}$ for our system. The numerical solution for the trajectories of $2^{21}$ particles requires huge computer memory and will be a time consuming task. Instead of taking the actual number of particles, we consider macroparticles~\cite{Braubook} in our simulations, where each macroparticle carries a charge larger than the charge on actual particle, but the same charge to mass ratio. In this way, we have taken $2^{17}$ particles, which carry the same charge as carried by the actual electron bunch, and can be easily handled in the numerical simulations. In order to simulate the flat electron beam, we have put all the electrons in one layer, which is located at a height $h$. Further, to model the width $\Delta y$ of the electron beam, this layer consists of $2^5$ arrays, which are equally distributed along the $y$-direction, in the range $(-\Delta y/2,\Delta y/2)$. Each array consists of $2^{12}$ electrons, which are propagating along the $z$-direction, and all electrons in one array will see the same magnitude of the electric field, depending upon the array position along the $y$-direction. The energy and phase of electrons will evolve in accordance of Eqs. (19) and (20), respectively. In each array, the electron beam is initialised in the phase space by using the quiet start scheme. In this scheme, electrons are assumed to have a uniform distribution in the phase space, where the phase of the $i$th electron is set to be $2\pi i/N$. Here, $N$ is the total number of electrons in an array. The initial dimensionless electric field in the system is set to be very small i.e., $\mathcal{E}$=0.001, and the input electron beam is considered to be monoenergetic with $\eta=2.6$.

\begin{figure}[t]
\includegraphics[width=13.0cm]{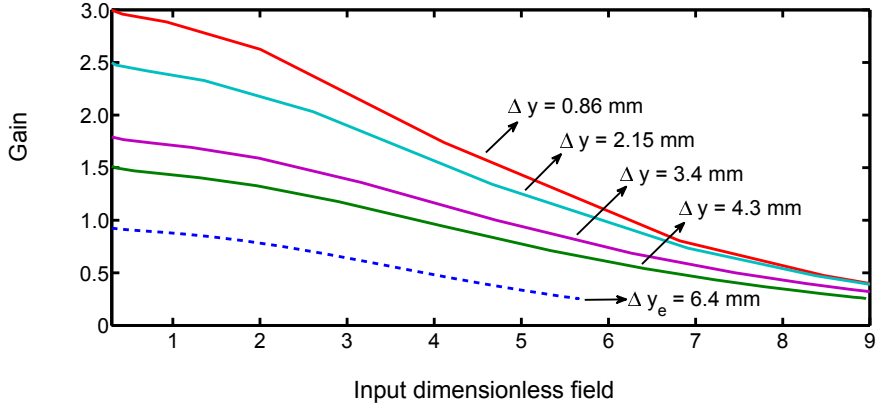}
\caption{Plot of net gain as a function of input electric field in a CFEL  driven by a monoenergetic flat electron beam. Dashed curve shows gain plot for a CFEL without any side wall, and solid curves represent the case of sidewall CFEL having sidewall spacing $w$=4.3 mm.}
\end{figure}

The main parameters used in the simulations have been listed in Table 1. In the CFEL system, gain has $e^{-2\Gamma h}$ dependency on the height $h$ of the flat beam, and therefore it is desirable that $h\leq 1/2\Gamma$ to have a sufficient beam-wave interaction, where $\Gamma=2\pi/\beta_R\gamma_R\lambda$. We take $h=1/2\Gamma=90.5~\mu$m in our calculations. For the metallic structure, we choose silver metal, and to minimize the effect of ohmic losses, we have kept the metallic structure at low temperature i.e., 77 K, which is the boiling point temperature of liquid nitrogen. At 77 K, the conductivity of silver metal is given by $3.3\times 10^8/\Omega$-m~\cite{Silverresitivity}, and by following the prescription given in Appendix~A, the ohmic attenuation coefficient is calculated as 1.9 per m. The value of tangent loss ($\tan \delta$) for GaAs dielectric at 77 K is $2\times 10^{-5}$~\cite{TangentGaAs}, and by using this value in Eq.~(A.16), we obtained the dielectric attenuation coefficient as 0.10 per m. It is to be noted that the dielectric losses are low as compared to the ohmic losses. The single pass loss ($1-e^{-2\alpha L}$) due to attenuation in the system is calculated as 18.4 $\%$. The sidewall spacing $w$ is taken as $w=(2/3)\sqrt{\pi\beta_p\lambda L/4}=4.3$ mm. In Fig. 3, gain has been plotted as a function of input field for different values of electron beam width, assuming an initially mono-energetic, flat electron beam. The dashed curve shows gain of a CFEL without any side wall, and solid curves represent gain of a side wall CFEL, for different values of electron beam width  $\Delta y$. The values of small-signal gain obtained by using the numerical simulations are in well agreement with the analytical results obtained using Eq.~(21), and these results have been summarized in Table 2. It is observed that in the case of the sidewall configuration of CFEL, we can obtain an enhanced gain; up to a factor of 3 as compared to the gain of CFEL without any side wall. For the sidewall configuration of CFEL, gain is enhanced when we decrease the electron beam width. It can be noted that as we decrease $\Delta y$, the agreement between the gain obtained using numerical simulations, and gain calculated using Eq.~(21) becomes better. For further simulations, we take electron beam size in the $y$-direction as $\Delta y=w/2=2.15$ mm.

\begin{table}[t]
\caption{\label{tab:table2}Comparison of analytical and simulation results for the net small-signal gain of CFEL, assuming different values of electron beam width. Here, $\Delta x=0$, and $w=4.3$ mm.}
\begin{ruledtabular}
\begin{tabular}{l c r}
 Electron beam width & Analytically calculated gain & Simulation results for gain\\
 \hline
 6.4 mm(without side walls) & 95 $\%$ & 93 $\%$\\
 4.3 mm & 115 $\%$ & 150 $\%$\\
 3.4 mm & 170 $\%$ & 179 $\%$\\
 2.15 mm & 248 $\%$ & 250 $\%$\\
 0.86 mm & 299 $\%$ & 300 $\%$\\
\end{tabular}
\end{ruledtabular}
\end{table}
\begin{figure}[t]
\includegraphics[width=13.0cm]{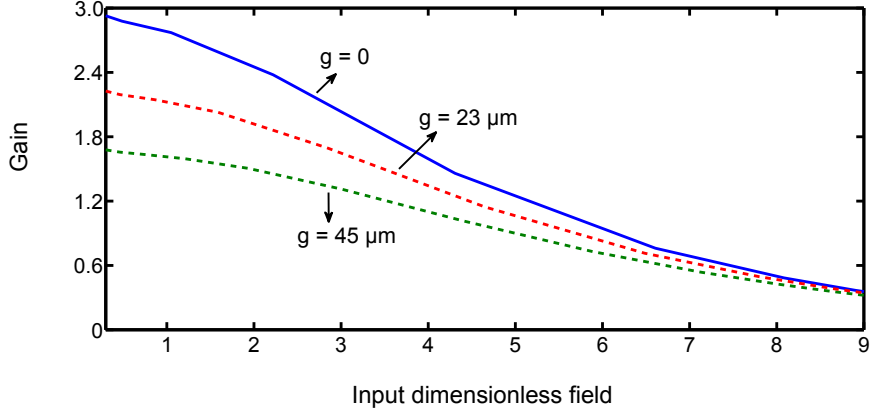}
\caption{Plot of net gain as a function of input electric field for different values of gap $g$ between the lower edge of the thick electron beam, and the dielectric surface in a sidewall CFEL. The electron beam is monoenergetic with $\Delta x$=181~$\mu$m, and $\Delta y=2.15$ mm.}
\end{figure}

Next, we take into account the effect of finite beam-thickness in the vertical direction in simulations by following the prescription given in Appendix~B. We first consider an electron beam of thickness $\Delta x=2h=181~\mu$m, with its centroid located at $x=h$. The gap $g$ between the lower edge of the electron beam and the dielectric surface is taken as zero. The finite-thickness of the electron beam is represented by $N_l$  number of layers distributed along the vertical direction. In our simulations, we have taken $N_l=4$. We have solved coupled Maxwell-Lorentz equations [Eqs.~(B.2-B4)] for this case, and plotted the results for the gain as a function of input dimensionless field, as shown by solid curve in Fig.~4. The small-signal gain is obtained as 293 $\%$ for the thick beam case, which is about 17.2 $\%$ higher compare to the case of a flat beam. This is because for a thick electron beam, the enhancement in gain due to some electrons coming closer to the dielectric surface is significant due to $e^{-2\Gamma h}$ dependency of the gain. These results are in agreement with the analytical formula for the gain given by Eq.~(B.5). We have also varied the gap $g$ between the lower edge of the electron beam and the dielectric surface. The CFEL gain for $g=23~\mu$m, and $45~\mu$m are shown in Fig. 4. The vertical thickness of the electron beam is taken as 181 $\mu$m. As expected, the gain decreases exponentially with the increase in the gap $g$. We would like to mention that we have also performed the numerical simulations for a thick beam by taking $N_l=2,8$ and 12, and observed that for $N_l\geq 4$, the simulation results converge. Most importantly, the converged result for the gain is in good agreement with the gain calculated using Eq.~(B.5). For further simulations discussed in this paper, we have considered $g=0~\mu$m, and taken $N_l=4$ to represent the electron beam with vertical thickness of 181 $\mu$m. 

\begin{figure}[t]
\includegraphics[width=13.0cm]{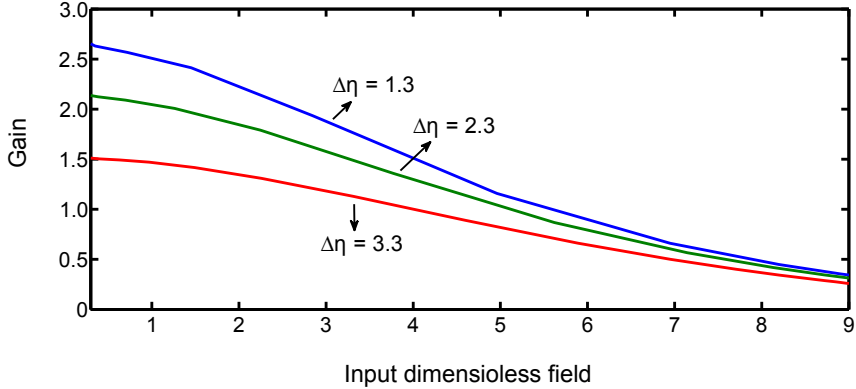}
\caption{Plot of net gain as a function of input electric field in the sidewall CFEL driven by a thick electron beam having finite energy spread.}
\end{figure}

We now consider the effect of energy spread in the electron beam. For this, we assume uniform distribution of electron energy around the mean value. We have considered three cases corresponding to relative rms energy spread of 0.6 $\%$, 1.0 $\%$, and 1.5 $\%$; which corresponds to half width in $\Delta\eta$ as 1.3, 2.3, and 3.3, respectively. Note that $\Delta\eta=k_zL\Delta\gamma/\beta_R^2\gamma_R^3$. As shown in Fig.~5, the small-signal gain is reduced by 10 $\%$ for an electron beam having 0.6 $\%$ relative rms energy spread, as compared to the case of monoenergetic electron beam. We find that as the energy spread increases, the small-signal gain of the system decreases. For further simulations discussed in the paper, we have considered  0.6 $\%$ relative rms energy spread. We would like to mention that as in conventional FELs, the effect of transverse emittance can be considered in terms of equivalent energy spread~\cite{Braubook}. The equivalent rms energy spread corresponding a normalized emittance $\bm{\varepsilon_{n,rms}}$ is given by $\gamma_R mc^2\bm{\varepsilon_{n,rms}}^2/2\sigma_e^2$, where $\sigma_e$ is the rms beam size. Since the emittance of the electron beam considered in our analysis is very small, the equivalent relative rms energy spread is less than 0.1 \% , and therefore not significant. However, it is important to note that we need very low emittance, such that a small beam size can be maintained over the interaction length, as will be discussed later in this section. This condition is particularly important for the vertical direction, since we need to maintain a very small beam size in the vertical direction. Larger value of vertical emittance will lead to larger beam size in the vertical direction, resulting in an increase in the height of the beam centroid above the dielectric surface. This will deteriorate the value of gain, and will lead to a $e^{-4\Gamma\sqrt{\bm{\varepsilon}_xL/\beta_R\gamma_R}}$ type dependence of gain on the normalized rms vertical emittance $\bm{\varepsilon}_x$~\cite{CFEL}.

The CFEL system discussed here is a low gain system, and to achieve saturation, the device has to be operated in the oscillator configuration. For this purpose, a set of mirrors is used to provide an external feedback. One mirror is assumed to have 100 $\%$ reflectivity for the field amplitude, while the second mirror is assumed to have the reflectivity of field amplitude as 98 $\%$. A fraction of the intra-cavity power can be out-coupled through the second mirror at the downstream end, and the output THz radiation can be guided via suitable optical arrangements to the nearby experimental station. In this configuration, the portion of the electromagnetic field that is reflected from the second mirror at the downstream end is then propagated to the upstream mirror (which is 100 \% reflective), and becomes input field for the next pass. It is to be noted that the electromagnetic field will be attenuated due to dielectric and ohmic losses as it propagates from the beginning to the end of the dielectric slab, and also during its back propagation from the end point to the beginning of the dielectric slab. The power in the surface mode decays by a factor of $e^{-2\alpha L}$ during the propagation of the field from the end point to the beginning of the dielectric slab. The input field for the next pass is $e^{-\alpha L}$ times the field, which is reflected from the downstream mirror in the previous pass. The coupled Maxwell-Lorentz equations have been solved in the view of the above mentioned conditions for the oscillator configuration, to obtain the power in the surface mode.

\begin{figure}[t]
\includegraphics[width=13.0cm]{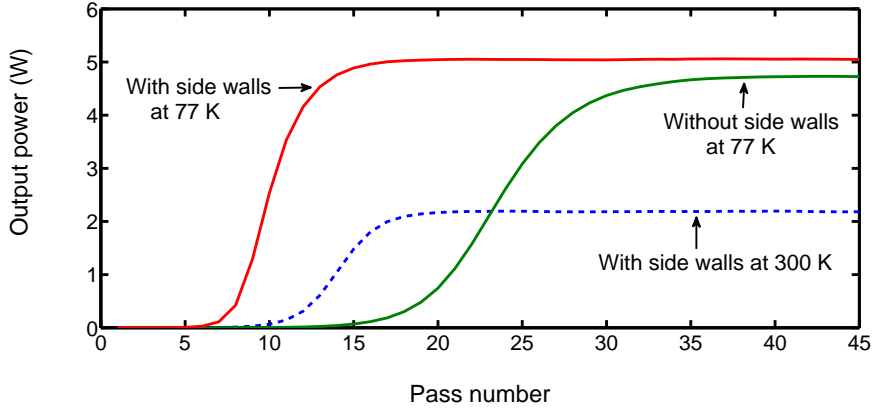}
\caption{Plot of output power as a function of pass number for a sidewall CFEL, and for a CFEL without any side wall. The solid curves represent the case where the dielectric, and the metallic structure are kept at 77 K, whereas the dashed curve shows output power for the case having dielectric, and metallic structure at 300 K.}
\end{figure}

The solid curves in Fig.~6 show output power as a function of pass number for the CFEL with and without side walls, where the temperature of the metallic, and the dielectric structure is taken as 77 K. The sidewall CFEL as compared to the CFEL without any side wall, saturates early, at around 20th pass. The output power at the saturation of the sidewall CFEL is 5.1 W, which is higher compared to the case of CFEL without any side wall. The input electron beam power is given by $P_b=1.6$ kW, and efficiency of the sidewall CFEL comes out to be 0.32 $\%$. A rough estimate for the upper bound of the efficiency of a CFEL system can be given by~\cite{Walsh2}
\begin{eqnarray}
\eta=\frac{\beta_R^3\gamma_R^3}{(\gamma_R-1)}\frac{\lambda}{L}.
\end{eqnarray}  
The above formula gives us the fraction of electron beam energy, which appears in the form of outcoupled power, plus the heat dissipated in the system.  In the oscillator configuration, the power which is outcoupled through the outcoupling mirror is given by $P_{in}(1-R^2)$, where $P_{in}$ is the mean intra-cavity power, and $R=0.98$ is the reflection coefficient of the out-coupling mirror. Taking the effect of attenuation, there will be a round trip loss of power as given by $P_{in}(1-e^{-4\alpha L})$. Considering these effects, efficiency of the system is understood as  $\eta=P_{in}((1-R^2)+(1-e^{-4\alpha L}))/P_b$. We define $\eta_{sim}$ as the efficiency observed in the simulation, which is also the actual efficiency, and represents the fraction of electron beam power that appears in the form of outcoupled power, i.e.,  $\eta_{sim}=P_{in}((1-R^2))/P_b$. This gives us $\eta_{sim}=[(1-R^2)/((1-R^2)+(1-e^{-4\alpha L}))]\eta$. Thus, the upper bound for $\eta_{sim}$ is obtained by multiplying a factor of $(1-R^2)/((1-R^2)+(1-e^{-4\alpha L}))$ in Eq.~(23). For the prescribed parameters, we find an upper bound of $\eta_{sim}$ for the sidewall CFEL as 0.51 $\%$, which is in agreement with the results of our numerical simulations.

In Fig.~6, we have also plotted the output power of a sidewall CFEL with its dielectric, and metallic structure kept at room temperature i.e., 300 K. At 300 K, the conductivity of silver is $6.3\times 10^7/\Omega$-m~\cite{Silverresitivity}, and the tangent loss for the GaAs is $2\times 10^{-4}$~\cite{TangentGaAs}. The total attenuation coefficient is calculated as $\alpha=5.5$ per m at 300 K, which results in round trip loss (in power) of 66.7 $\%$. The system is above the threshold at 300 K, and saturates at output power of 2.2 W, as shown by the dashed curve in Fig.~6. The efficiency of the CFEL system is 0.14 $\%$ at room temperature, which is around 56 $\%$ less than the efficiency of the CFEL system at 77 K. This is because at room temperature, the attenuation of the surface mode increases, and reduces the output power of the CFEL. One has to reduce the losses to obtain an optimum performance of the CFEL system.

\begin{figure}[t]
\includegraphics[width=13.0cm]{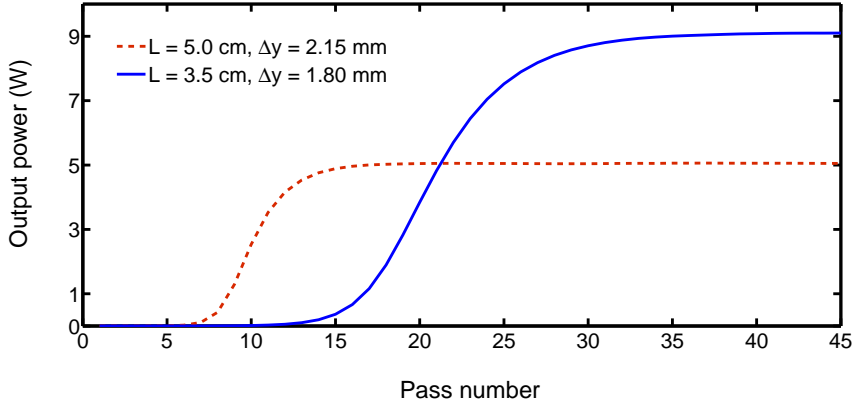}
\caption{Plot of output power as a function of pass number in sidewall CFEL with its metallic, and dielectric structure kept at 77 K. The dashed and solid curves represent the power of a side wall CFEL with $L=5$ cm, and $L=3.5$ cm, respectively.}
\end{figure}

Next, we would like to discuss that for the case of guided CFELs, we can take a shorter interaction length to improve the performance of the system. As mentioned earlier, by taking smaller electron beam size $\Delta y$, one can achieve a higher value of gain. However, the condition on the electron beam horizontal emittance becomes stringent at the smaller beam size. In order to maintain a given value of $\Delta y$ over the interaction length $L$, we require that the normalized rms beam emittance should satisfy $\bm{\varepsilon_y}\leq \beta_R\gamma_R\Delta y^2/16L$~\cite{KimPRSTB}. Similar condition can be written with subscript $x$ in the $x$-direction. For $L=5$ cm, we have taken $\Delta y=w/2=2.15$ mm, and $\Delta x=181~\mu$m, which require a flat beam with fine beam emittances $\bm{\varepsilon_y}\leq 2.5\times 10^{-6}$ m-rad, and $\bm{\varepsilon_x}\leq 1.8\times 10^{-8}$ m-rad. Hence, to drive the CFEL system, we require an asymmetric electron beam with transverse emittance ratio $\bm{\varepsilon_y}/\bm{\varepsilon_x}\sim 139$. Such a DC electron beam can be produced by employing a round to flat beam transformation to the electron beam produced using an initially round, thermionic cathode such as LaB$_6$, as described in Ref.~\cite{KimPRSTB}. This technique of round to flat beam transformation has been demonstrated experimentally at Fermi National Accelerator Laboratory to generate an electron beam directly from a photoinjector with transverse emittance ratio of 100~\cite{FlatphaseE1,Flatphase3}. The stringent requirements on the beam emittances can be relaxed by taking a smaller interaction length $L$. The additional advantage is that we can achieve high efficiency at small $L$, as it is clear from Eq.~(23). It is to be noted that with decrease in $L$, the gain will decrease, and one has to optimize $L$ such that the system can overcome the losses. This criterion can be easily achieved in the sidewall CFEL, which has improved gain as compared to the CFEL without side walls. We take $L$=3.5 cm for a sidewall CFEL, for which the net small-signal gain is 94.2 $\%$. We require an electron beam with relaxed transverse emittance ratio of 100 for this case. In Fig.~7, the solid curve shows the output power of a sidewall CFEL with $L=3.5$ cm. As expected, CFEL at shorter interaction length saturates at higher power of about 9.1 W, and has an increased efficiency of 0.6 $\%$. Analysis in the linear as well as in the non-linear regime shows that the guiding of surface mode in the CFEL is helpful in improving the performance of the system.

\section{\label{sec:level3}Discussions and conclusions}
In this paper, we have presented a theory of waveguided $\check{\text{C}}$erenkov FEL in the single slab geometry, where the waveguiding is provided by metallic side walls. The waveguiding is particularly useful to enhance the performance of a CFEL at longer wavelength, where diffraction effects are prominent and reduce the gain. Due to waveguiding, we can reduce the mode width below the value which can be achieved in the absence of waveguiding. The waveguided surface modes have been studied earlier for the undulator based FELs~\cite{WaveguideFEL}, Smith-Purcell FELs~\cite{SWLi,SWAJB,SWDonohue}, and for the cylindrical~\cite{Walsh1,FreundGanguly,CiocciCimento}, and rectangular~\cite{DingPlasma,DingChina} geometry of CFELs. In this paper, we have extended this approach for the single slab based open geometry of CFELs.

Our analysis is built on the model discussed earlier in the Refs.~\cite{ColsonFEL1,LevushBWO}. We have set up the coupled Maxwell-Lorentz equations for the sidewall CFEL in single slab configuration. These equations have been solved analytically to obtain an approximate solution for the small-signal gain, and the growth rate of the system. In the non-linear regime, we solved these equations numerically by using the Leapfrog scheme to obtain the output power at saturation, and included the effects due to finite energy spread, and finite beam size. We find that a CFEL with length $L=3.5$ cm, and spacing between the side walls as 3.6 mm, can be operated at 0.11 THz by using a 47.5 keV, 35 mA electron beam to give output power of almost 10 W with an efficiency of 0.6 $\%$ at saturation. The relative rms energy spread is taken as 0.6 $\%$ for the electron beam. We also worked out the requirements on the electron beam emittance for this case.

It is important to note here that while calculating the output power, we have assumed that a fraction of the total intra-cavity power can be outcoupled through the outcoupling mirror. Total intra-cavity power is the sum of power in radiative mode inside the dielectric slab, and the power in evanescent mode in the vacuum region. Although the radiation inside the dielectric may undergo total internal reflection at the ends, the power in the evanescent mode in the vacuum region can be outcoupled through a hole in the mirror, where it will get converted to useful radiative mode. Here, the hole can also be used to extract the electron beam. Although the detailed analysis of outcoupling will be an involved one; we have assumed that with a suitable design of outcoupling system, a small fraction (4 $\%$ in our case) of the intra-cavity power can be outcoupled through the downstream mirror. In order to model this situation in a simple manner, we have assumed that the downstream mirror is semi-transparent (with 98 $\%$ reflectivity in the field amplitude).  The upstream mirror in our analysis is assumed to be transparent to the electron beam. In practice, the electron beam is injected by putting a hole in the upstream mirror, and it is not 100 $\%$ reflective, as assumed in our calculations. Also, due to the presence of hole in the mirrors, field inside the cavity will contain higher order modes, which are not considered in our model. These are some of the approximations used in our analysis to model the oscillator configuration. 

We would like to mention that for the beam parameters considered in our analysis, the collective effect due to space charge is not very important. This is because the number of plasma oscillations performed as the beam passes through the interaction region is significantly less than one~\cite{Braubook}. In our analysis, we have not considered the time-dependent effects while solving the coupled Maxwell-Lorentz equations. This is because we have used a DC electron beam to drive the CFEL system. A CFEL can also be operated by using a bunched electron beam produced by an rf accelerator~\cite{EXPCFEL2,EXPCFEL90}, and the time dependent effects i.e., slippage and anti-lethargy effects become prominent for very short electron pulses, and at long wavelengths~\cite{pulseCFEL}.

We have not considered the fluctuations in the dielectric slab properties (relative permittivity, slab uniformity), which can reduce the gain as well as the saturated power~\cite{fuente1,fuente2}. We can however estimate the tolerable fluctuations of liner thickness and relative permittivity along the length of the dielectric. For a given beam energy, if we vary the liner thickness $d$, or relative permittivity $\epsilon$, the resonant wavelength will change along the length. Like in the case of conventional FEL, this will result in a spread in the detuning parameter $\eta$, and therefore gain will decrease. In order to ensure that the gain does not deteriorate significantly, it is required that the spread in $\eta$ is much less than 2$\pi$. Using this criterion, we have estimated that the tolerable fluctuation in liner thickness and relative permittivity are around 10 $\mu$m, and 5\% respectively, for the CFEL parameters discussed in this paper.

We want to emphasize that in Ref.~\cite{CFEL}, we performed the analysis of CFEL having no side walls, by analysing the singularity in the reflectivity of the dielectric slab (placed on a conducting surface) for the incident evanescent wave, and deriving the expressions for the parameters $\chi$ and $\chi_1$. The growth of surface mode is understood in terms of the $\chi$ parameter, and the ac space charge effect is studied by using the parameter $\chi_1$. In Refs.~\cite{CFEL,VinitPRE}, it has been shown that the above mentioned approach, and the approach followed by Levush $et~al.$~\cite{LevushBWO}, which is also adopted in the present paper; give same result for the case of CFELs, and SP-FELs having no side walls. We have checked that even in the presence of side walls, the $\chi$ and $\chi_1$ parameters can be derived and an analysis can be performed in terms of these parameters, and it gives the same results as described in this paper. The same will be applicable in the SP-FELs~\cite{VinitPRE,VinitFEL05}, and the approach based on Maxwell-Lorentz equations can certainly be extended to the case of sidewall SP-FELs. We would like to mention that our analysis can also be extended to study the CFEL based on negative refractive index material. CFEL based on the negative index material will work like a BWO~\cite{Li3}, and can be studied by following the approach given in Refs.~\cite{VinitPRE,KimPRSTB}.

To conclude, we have proposed a CFEL with metallic side walls. A detailed analysis for the guided surface mode is presented. Realistic effects such as effects due to transverse variation of the field, finite beam-size, attenuation effects, and effects due to finite energy spread have been included in the analysis. Our analysis shows that a sidewall CFEL has higher gain as compared to the CFEL without side walls, and can give significant output power in the THz regime, even after including the attenuation effects, and effects due to finite energy spread. The presented analysis can be very useful to design a practical CFEL device with side walls.

\section*{Acknowledgements}
We are grateful to an anonymous referee for several helpful comments and suggestions. It is a pleasure to thank Professor Arup Banerjee for useful discussions. We acknowledge Professor S. B. Roy and Professor P. D. Gupta for constant encouragement. One of us (YK) gratefully acknowledges Homi Bhabha National Institute, Department of Atomic Energy (India) for financial support during the research work.

\appendix
\section{POWER, GROUP VELOCITY AND ATTENUATION COEFFICIENT OF THE SURFACE MODE SUPPORTED IN A SIDEWALL $\check{\text{C}}$ERENKOV FEL}

In this Appendix, we have evaluated power and group velocity of the surface mode supported in a CFEL with side walls. The obtained results are then used to calculate the attenuation coefficient of the surface mode due to the dielectric, and ohmic losses present in the system. The schematic of the sidewall CFEL is shown in Fig.~1, where the metallic side walls are placed at $y=\pm w/2$. The electromagnetic surface mode supported by this structure can be obtained by combining two plane evanescent waves, each having frequency $\omega$ and wave vector $k_0=\sqrt{k_y^2+k_z^2}$, travelling in different directions in the ($y,z$) plane. The dielectric slab is an isotropic structure in the ($y,z$) plane, and the optical properties of the system will remain invariant under any arbitrary rotation in the ($y,z$) plane. By using the property of isotropy in the results derived earlier for a CFEL without any side wall~\cite{CFEL}, we obtain the electromagnetic field components of the surface mode in the vacuum region of the sidewall CFEL as:

\begin{eqnarray}
H_y^I&=&(k_zH/k_0)\cos(k_yy)^{i(k_zz-\omega t)}e^{-\Gamma (x-h)}+\text{c.c.,} ~~~~~~~~~~~~~~~~~~~~~~~~~~~~\\
H_z^I&=&(-ik_yH/k_0)\sin(k_yy)e^{i(k_zz-\omega t)}e^{-\Gamma (x-h)}+\text{c.c.,}\\
E_x^I&=&(k_0H/\epsilon_0\omega)\cos(k_yy)e^{i(k_zz-\omega t)}e^{-\Gamma (x-h)}+\text{c.c.,}\\
E_y^I&=&(k_y\Gamma H/\epsilon_0\omega k_0)\sin(k_yy)e^{i(k_zz-\omega t)}e^{-\Gamma (x-h)}+\text{c.c.,}\\
E_z^I&=&(-ik_z\Gamma H/\epsilon_0\omega k_0)\cos(k_yy)e^{i(k_zz-\omega t)}e^{-\Gamma (x-h)}+\text{c.c.,}
\end{eqnarray}
and $H_x^I=0$. Here, $k_z=2\pi/\beta\lambda$, $k_y=\pi/w$, and $\Gamma=\sqrt{k_y^2+k_z^2-\omega^2/c^2}$. Inside the dielectric medium, the components of electromagnetic field are given by

\begin{eqnarray}
H_y^{II}&=&\frac{\epsilon k_z\Gamma H}{k_1k_0}\frac{\cos[k_1(x+d)]}{\sin(k_1d)}\cos(k_yy)e^{\Gamma h}e^{i(k_zz-\omega t)}+\text{c.c.,} 
\end{eqnarray}
\begin{eqnarray}
H_z^{II}&=&\frac{-i\epsilon k_y \Gamma H}{k_1k_0}\frac{\cos[k_1(x+d)]}{\sin(k_1d)}\sin(k_yy)e^{\Gamma h}e^{i(k_zz-\omega t)}+\text{c.c.,}
\end{eqnarray}
\begin{eqnarray}
E_x^{II}&=&\frac{k_0\Gamma H}{\epsilon_0\omega k_1}\frac{\cos[k_1(x+d)]}{\sin(k_1d)}\cos(k_yy)e^{\Gamma h}e^{i(k_zz-\omega t)}+\text{c.c.,}
\end{eqnarray}
\begin{eqnarray}
E_y^{II}&=&\frac{k_y\Gamma H}{\epsilon_0\omega k_0}\frac{\sin[k_1(x+d)]}{\sin(k_1d)}\sin(k_yy)e^{\Gamma h}e^{i(k_zz-\omega t)}+\text{c.c.,}
\end{eqnarray}
\begin{eqnarray}
E_z^{II}&=&\frac{-ik_z\Gamma H}{\epsilon_0\omega k_0}\frac{\sin[k_1(x+d)]}{\sin(k_1d)}\cos(k_yy)e^{\Gamma h}e^{i(k_zz-\omega t)}+\text{c.c..}
\end{eqnarray}
Here, $H_x^{II}=0$ and $k_1=\sqrt{\epsilon\omega^2/c^2-k_y^2-k_z^2}$. Power flow in the surface mode can be evaluated by integrating the Poynting vector over $x$ in the range ($-d,\infty$), and over $y$ in the range ($-w/2,w/2$). Total power in the surface mode is sum of power flow in the vacuum and inside the dielectric medium, which we obtain as:
\begin{eqnarray}
P=\frac{w\beta_p\gamma_p^3}{2k_zZ_0}\bigg[1+\frac{1}{\epsilon^2a^2}+\frac{k_0d(1+a^2)}{\epsilon \gamma_p a^2}\bigg]E_z^2e^{2\Gamma h}.
\end{eqnarray}
Here, $\beta_p=\omega/ck_0$ is the phase velocity of the surface mode in unit of $c$, $\gamma_p=1/\sqrt{1-\beta_p^2}$, $a=(\gamma_p/\epsilon)\sqrt{\epsilon\beta_p^2-1}$, and $Z_0=1/\epsilon_0c$ is the characteristic impedance of the free space. Note that above equation has been expressed in term of half amplitude $E_z$ of the peak field at the location of electron beam by using Eq.~(A.5). It should be noted that the total power in the sidewall CFEL is $k_z/2k_0$ times the power contained in the surface mode supported in the CFEL without any side wall~\cite{CFEL}. This is obvious as the electromagnetic fields in the sidewall CFEL are propagating at an angle, whose cosine gives the factor $k_z/k_0$, with respect to the $z$-direction. The factor $1/2$ accounts for the variation of electromagnetic fields along the $y$-direction. The energy stored in the fields is obtained by integrating the energy density over the volume of the dielectric and over the volume of the vacuum region. The expression for the energy stored per unit mode width $w$ per unit length in the $z$-direction is obtained as:
\begin{eqnarray}
\frac{\mathcal{U}}{E_z^2}=\frac{k_o\gamma_p^3}{2ck_z^2Z_0}\bigg[1+\frac{1}{\epsilon^2a^2}+\frac{k_0d\beta_p^2(1+a^2)}{\gamma_p a^2}\bigg]e^{2\Gamma h}.
\end{eqnarray}
The energy velocity of the electromagnetic fields is given by $P/w\mathcal{U}$. For $\check{\text{C}}$erenkov FEL, the energy velocity is equal to the group velocity~\cite{CFEL}. Using Eqs.(A.11) and (A.12), we obtain the group velocity of the surface mode supported in the sidewall CFEL as:
\begin{equation}
v_g=\frac{\beta_pck_z}{k_0}\frac{\big[\beta_p^2\gamma_p^3(\epsilon-1)+k_0d\epsilon(1+a^2)\big]}{\big[\beta_p^2\gamma_p^3(\epsilon-1)+k_0d\epsilon^2\beta_p^2(1+a^2)\big]}~.
\end{equation}  
In the sidewall CFEL, the group velocity of the surface mode is $k_z/k_0$ times the group velocity of the CFEL without any side wall~\cite{CFEL}.

Next, we evaluate the attenuation coefficient of the surface mode, which is given by~\cite{Liao}
\begin{eqnarray}
\label{eq:attenuation1}
\alpha^{m,d}=\frac{P^{m,d}_l}{2P},
\end{eqnarray}
where $P_l$ represents the power loss per unit length along the $z$-direction, and the subscripts $m$ and $d$ are used to represent the metallic conductor and the dielectric medium, respectively. In the metallic structure with finite conductivity $\sigma$, dissipation of power occurs due to the ohmic losses. At the location $x=-d$, the power loss per unit length along the surface of metal is given by:$(R_s/2)\int_{-w/2}^{w/2} \vert H_y^{II}\vert^2 dy$~\cite{Jackson}, where $R_s=\sqrt{\mu_0\omega/2\sigma}$ is surface resistance of the metal. At the location of side walls, i.e., at $y=\pm w/2$, we can write the power loss per unit length as: $(R_s/2)(\int_{-d}^{0} \vert H_z^{II}\vert^2 dx+\int_0^{\infty} \vert H_z^I\vert^2 dx)$~\cite{Jackson}. Note that the limit of integration over $x$ has been extended up to infinity, which is due to the fact that the electromagnetic field is decaying in the $x$-direction, and has a negligible value at the top of the side walls. The total power dissipated along the metallic surface $P_l^m$ is sum of power dissipated at the metallic surface located at $x=-d$, and the power dissipated at the side walls. By performing the required algebra for $P_l^m$ and by using the expression (A.14), we obtain the ohmic attenuation coefficient as:

\begin{eqnarray}
\hspace*{-15pt}\alpha^m=\frac{2\beta_pR_sk_y^2}{\gamma_p wk_z\Gamma Z_0}\frac{\big[1+(w\Gamma k_z^2/2a^2k_y^2)(1+a^2)+(\epsilon^2\Gamma^3/k_1^3)(a+k_1d(1+a^2))\big]}{\big[1+(1/\epsilon^2a^2)+(k_0d/\epsilon\gamma_pa^2)(1+a^2)\big]}.~~~~
\end{eqnarray} 

Inside the dielectric medium, losses are described in the terms of complex relative permittivity $\tilde{\epsilon}=\epsilon-i\epsilon''$ with tangent loss defined as $\tan\delta=\epsilon''/\epsilon$~\cite{dielectricloss1}. Now, power loss per unit length inside the dielectric material is given by ~$P^{d}_l=\epsilon_0\epsilon\omega\tan\delta\int_{-d}^0\int_{-w/2}^{w/2}\big(\vert E^{II}_x\vert^2+\vert E^{II}_y\vert^2+\vert E^{II}_z\vert^2\big)dxdy$. By using Eqs.~(A.8-A.10), we first evaluate $P_l^d$, and then by using Eq.~(A.14), we obtain the attenuation coefficient $\alpha^d$ as:
\begin{eqnarray}
\label{eq:alpha2}
\alpha^{d}=\frac{k_0^2\tan\delta}{2k_z}\frac{[\gamma_p(2-\epsilon\beta_p^2)+\epsilon^2\beta_p^2k_0d(1+a^2)]}{[\gamma_p(1+\epsilon^2a^2)+\epsilon k_0d(1+a^2)]}.
\end{eqnarray}
Total attenuation coefficient $\alpha$ of the surface wave can be written as $\alpha=\alpha^m+\alpha^d$, which gives attenuation due to both the ohmic loss and the loss presents in the dielectric.

\section{MAXWELL-LORENTZ EQUATIONS AND SMALL-SIGNAL GAIN FOR FINITE-THICKNESS BEAM}
Here, we extend the coupled Maxwell-Lorentz equations described in Sec. 2.2 to include the finite-thickness $\Delta x$ of the electron beam in the $x$-direction. We assume that the thick electron beam can be described as a combination of $N_l$ layers, where the $l$th layer is at a height $h_l=(2l-1)\Delta x/2N_l$ with current $I_l$, and thickness $\Delta x_l=\Delta x/N_l$, corresponding to $x$ in the range $((l-1)\Delta x/N_l,l\Delta x/N_l )$. The dimensionless current $\mathcal{J}_l$ for the $l$th layer is defined as:
\begin{eqnarray}
\mathcal{J}_l= \frac{4\pi k_zL^3}{Z_0\beta_R^2\gamma_R^3}\frac{I_l}{I_A}\frac{E_z^2}{v_g w\mathcal{U}}.
\end{eqnarray}
The electromagnetic surface mode interacts with current in all layers, and the Maxwell field equation can be written as: 
\begin{eqnarray}
 \frac{\partial \mathcal{E}}{\partial \xi}+\frac{\partial \mathcal{E}}{\partial \tau}= -\sum\limits_{l=1}^{l=N_l}\mathcal{J}_le^{-\Gamma h_l}\langle\cos(k_yy) e^{-i\psi} \rangle_l-\alpha L \mathcal{E},~~~~~~
\end{eqnarray}
where $\langle \cdots \rangle_l$ indicates averaging over all the electrons present in the $l$th layer. The Lorentz equations for the energy and phase of $i$th electron in $l$th layer are given by:
\begin{equation}
\frac{\partial \eta_i^l}{\partial \xi}=\mathcal{E}\cos(k_yy)e^{i\psi_i}e^{-\Gamma h_l}+\text{c.c.},
\end{equation}
\begin{equation}
\frac{\partial \psi_i^l}{\partial \xi}=\eta_i^l.~~~~~~~~
\end{equation}
Equations (B.2-B.4) have to be solved numerically to study the beam-wave interaction for the finite-thickness electron beam. To find an analytical expression for gain in this case, we proceed as follows. Gain of a CFEL driven by the flat electron beam propagating at height $h$ varies as $e^{-2\Gamma h}$, as discussed earlier. The finite-thickness electron beam with $\Delta x=2h$, and its centroid at $x=h$ can be equivalently represented by a set of infinite number of flat layers located between $x=0$, and $x=2h$; and the gain for the finite-thickness case can be obtained by averaging over all the layers between $x=0$, and $x=2h$. This gives us a factor of $(1/2h)\int\limits_{0}^{2h}e^{-2\Gamma x}dx=(1-e^{-4\Gamma h})/4\Gamma h$. This factor together with the term $e^{2\Gamma h}$ has to be multiplied in the formula for gain given by Eq.~(21), to account for the effect of finite beam thickness. We therefore obtain the formula for the small-signal gain as:
\begin{eqnarray}
G= 6.75\times 10^{-2} \frac{16\pi k_zL^3}{Z_0\beta_R^2\gamma_R^3}\frac{I}{I_A}\frac{E_z^2}{v_g w\mathcal{U}}\frac{e^{2\Gamma h}(1-e^{-4\Gamma h})}{4\Gamma h}\bigg(\frac{\sin(k_y\Delta y/2)}{k_y\Delta y/2}\bigg)^2.
\end{eqnarray} 
Note that in our analysis, the electron beam is assumed to be uniformly distributed in the region above the dielectric surface in the $x$-direction. One can also simulate an arbitrary profile of the electron beam by taking different surface current density in different layers. 


\end{document}